\documentclass[journal]{IEEEtran}
\IEEEoverridecommandlockouts
\usepackage{tabularx}
\usepackage{cite}
\usepackage{amsmath,amssymb,amsfonts}
\usepackage{algorithmic}
\usepackage{graphicx}
\usepackage[linesnumbered,ruled]{algorithm2e}
\usepackage{subcaption}
\usepackage{textcomp}
\usepackage[table,xcdraw]{xcolor}
\usepackage{xcolor}
\usepackage{siunitx}

\def\BibTeX{{\rm B\kern-.05em{\sc i\kern-.025em b}\kern-.08em
    T\kern-.1667em\lower.7ex\hbox{E}\kern-.125emX}}
\begin{document}

\title{On the Optimization of Model Aggregation for Federated Learning at the Network Edge}

\author{
\IEEEauthorblockN{Mengyao Li$^{1}$, Noah Ploch$^{2}$, Sebastian Troia$^1$, Carlo Spatocco$^{1}$, Wolfgang Kellerer$^2$, Guido Maier$^1$}

\IEEEauthorblockA{$^1$\textit{\textit{Politecnico di Milano, Dipartimento di Elettronica, Informazione e Bioingegneria (DEIB), Milan, Italy}} \\}
\IEEEauthorblockA{$^2$\textit{\textit{Technical University Munich, School of Computation, Information, and Technology, Munich, Germany}}\\}
}

\maketitle

\begin{abstract}

The rapid increase in connected devices has significantly intensified the computational and communication demands on modern telecommunication networks. To address these challenges, integrating advanced Machine Learning (ML) techniques like Federated Learning (FL) with emerging paradigms such as Multi-access Edge Computing (MEC) and Software-Defined Wide Area Networks (SD-WANs) is crucial. This paper introduces online resource management strategies specifically designed for FL model aggregation, utilizing intermediate aggregation at edge nodes. Our analysis highlights the benefits of incorporating edge aggregators to reduce network link congestion and maximize the potential of edge computing nodes. However, the risk of   network congestion persists. To mitigate this, we propose a novel aggregation approach that deploys an aggregator overlay network. We present an Integer Linear Programming (ILP) model and a heuristic algorithm to optimize the routing within this overlay network. Our solution demonstrates improved adaptability to network resource utilization, significantly reducing FL training round failure rates by up to 15\% while also alleviating cloud link congestion.

\end{abstract}


\section{Introduction}

The ever-increasing volume of data generated by diverse devices and applications presents a significant challenge to existing computer networks. This highlights the urgent requirement for scalable and adaptable networking solutions that extend beyond the existing cloud-centric infrastructure \cite{filali2020multi}. 

Let's consider a smart city with a multitude of Internet of Things (IoT) devices continually gathering data from sensors integrated into various infrastructure elements, including but not limited to traffic lights, environmental sensors, and surveillance cameras. Concurrently, applications operating on these devices analyze this data in real time to provide insights or take actions. These actions may range from optimizing traffic flow to identifying irregularities or augmenting public safety measures. The substantial volume of data generated by these devices and applications daily warrants attention. The conventional method of consolidating all data processing activities within a centralized cloud data center is increasingly deemed impractical for several compelling reasons, such as latency and bandwidth. In the first case, some applications require low-latency responses, e.g. real-time traffic management or emergency response systems. In the second, transmitting massive volumes of data from numerous IoT devices to a centralized cloud server strains the available bandwidth, leading to congestion and potential bottlenecks. 

Multi-access Edge Computing (MEC) emerges as a game-changer networking paradigm, strategically leveraging the computational and storage capabilities of end devices (clients) and edge servers. MEC aims at bringing Machine Learning (ML) model training closer to the point of data generation, facilitating data processing outside traditional cloud data centers. MEC uses computational and communication resources situated at the edge of the network, thereby establishing intermediary network resources between end devices, such as IoT devices, and the cloud \cite{mach2017mobile}. As such, it enables mobile devices to run highly demanding applications while meeting strict delay requirements.

With the proliferation of MEC-based applications, users' expectations of Wide Area Network (WAN) connectivity performance have evolved dramatically. Beyond mere connectivity, also unwavering reliability, agility, and exceptional performance should be expected on WAN infrastructure.
Software-Defined Wide Area Network (SD-WAN) comes into play as a transformative paradigm shift in enterprise and business networking. At its core, SD-WAN is a software-based solution that promises streamlined deployment, elevated connectivity, and centralized control. It empowers organizations to dynamically manage their WANs through virtualization technology. In short, SD-WAN stands out for its remarkable capability to swiftly establish overlay networks, often within several seconds. Through Virtual Private Network (VPN) tunnels, SD-WAN can dynamically create network overlays, enabling organizations to adapt quickly to changing network demands and configurations \cite{yang2019software}. The agility offered by SD-WAN in creating overlay networks is particularly advantageous in environments where flexibility and scalability are important. In scenarios where new branch offices or far-edge devices (e.g. environmental sensors) need to be connected promptly or where temporary network configurations are required, SD-WAN excels in providing rapid end-to-end connectivity deployment without the need for extensive manual configuration. However, it is important to acknowledge that the speed of deployment with SD-WAN may come at the expense of certain network guarantees, such as latency and bandwidth assurances. While SD-WAN can rapidly establish connections, the reliance on VPN tunnels may introduce variability in network performance and susceptibility to fluctuations in network conditions \cite{yalda2022survey}.


While a majority of current ML applications remain cloud-centric, the use of the cloud for ML introduces challenges such as constant data uploading, leading to core network congestion and unacceptable latency \cite{lim2020federated}. 


Federated Learning (FL) has gained increasing attention due to its ability to collaboratively train a global ML model in a distributed manner without compromising the privacy of data held by individual clients \cite{mcmahan2017communication}. FL operates as a form of distributed ML, wherein multiple data owners collectively contribute to the training of a global model, thereby avoiding the need to share their raw data. Within the FL network architecture, entities include clients end devices, aggregators, and servers. Client devices store their original data locally and refrain from exchanging or transferring it. Instead, each local learner leverages the datasets of other learners solely through the global model, which is shared by the aggregator, without direct access to privacy-sensitive data \cite{niknam2020federated}. The FL workflow involves each device's local data for training, and then transmitting the trained model to the server for aggregation. Finally, the server distributes the updated model to all clients to advance the collective learning objective \cite{zhang2021survey}.

Recently, the embedding of FL at the network edge has attracted considerable attention both in the Industry and Academia. This interest stems from the benefits offered by both MEC and SD-WAN, such as improved efficiency in communication resource usage and enhanced user privacy \cite{lim2020federated}. However, implementing FL at the network edge entails numerous challenges due to the heterogeneity of the environment as compared to traditional data centers. These challenges encompass various aspects. Firstly, edge-to-cloud communication may suffer from slower and unstable connections, necessitating robust solutions to mitigate potential latency and reliability issues. Additionally, ensuring the trustworthiness of edge nodes becomes paramount to safeguard the integrity of FL processes, given that edge devices may be more susceptible to security vulnerabilities \cite{lim2020federated}. Furthermore, addressing the heterogeneity inherent in network and computing resource specifications at the edge requires the implementation of intelligent resource management strategies \cite{trindade2022resource}. These strategies are crucial for optimizing FL performance while accommodating variations in edge device capabilities and network conditions.

\begin{figure}[htbp]
\vspace{-3mm}
\centering
\includegraphics[width=0.45\textwidth,]{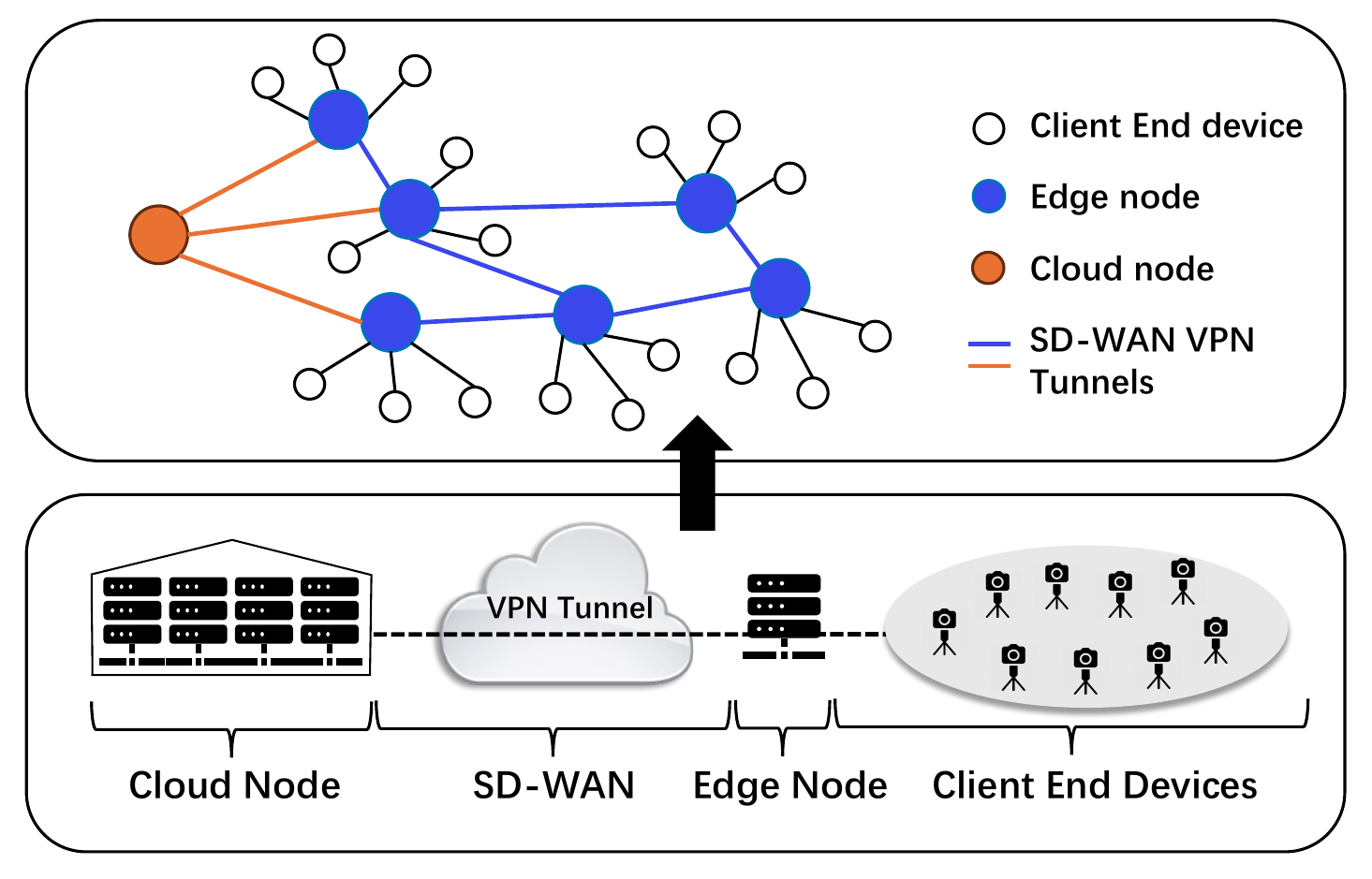}
\caption{Architecture of FL with MEC enhanced by SD-WAN.}
\label{fig:network}
\vspace{-3mm}
\end{figure}

In this work, we explore the deployment of FL within a MEC network enhanced by SD-WAN technology, with a particular focus on resource management during the FL training phase, as illustrated in Fig.\ref{fig:network}. Unlike conventional FL algorithms that primarily focus on aggregating local models while emphasizing privacy preservation and low latency, our proposed approach introduces a novel perspective by addressing the unique challenges posed by SD-WAN-enabled MEC environments. Existing FL algorithms typically do not consider the complexities of network connectivity, which is important in our scenario. Our approach takes into account the constraints of WAN connectivity, which is managed at the edge by SD-WAN, alongside the stringent limitations of both computational resources and network capacity. To optimize resource utilization, we perform aggregation of local models directly on strategically selected edge nodes, thereby significantly reducing tunnel capacity costs associated with cloud transmissions. To thoroughly evaluate and compare various aggregation strategies, we have developed a Discrete-Event Simulator (DES) capable of accurately modeling resource consumption behaviors in FL applications under different aggregation algorithms.

This work is supported by the WatchEDGE \cite{MAIER2024110248} project, whose goal is to study an advanced edge computing architecture distributed over several geographically-distant sites. Each site represents an “island” equipped with edge-computing capabilities (in-network storage and processing). The islands are interconnected to each other exploiting WAN connectivity controlled at the edge by an SD-WAN. The project addresses the challenge of providing an overall orchestration of applications and resources in the WatchEDGE infrastructure in a fully distributed way. Environmental surveillance and wildlife protection serve as use case for the proposed architecture, inducing the need for AI-based image processing on heterogeneous devices, see Fig.\ref{fig:network}.

The contribution of this paper is two-folded:
\begin{itemize}
    \item We propose an Integer Linear Programming (ILP) formulation of the edge-to-cloud FL model aggregation problem. Then, we propose a heuristic algorithm called Hierarchical Federated Edge Learning Mesh (HFEL-MESH) in order to cope with the scalability problem posed by the ILP. To validate the efficacy of our proposed methods, we conduct a thorough comparative analysis against a state-of-the-art algorithm as referenced by \cite{Liu:2020}.   
    \item We have developed a Python-based Discrete Event Simulator (DES) tailored specifically for the WatchEDGE project. This simulator is customized to emulate the infrastructure and applications inherent to the project's ecosystem, with a specific focus on AI algorithms. At its core, the simulator aims at replicating the complexity of the WatchEDGE infrastructure while facilitating the simulation of AI-based applications, particularly those related to FL for image recognition tasks at the edge.
\end{itemize}

The rest of the paper is organized as follows. Sec.~\ref{sec:related_work} discusses the related work within FL and edge networking/computing topics. Sec.~\ref{sec:FLRS} introduces the problem statement, the ILP model, and a novel scalable heuristic algorithm. Sec. \ref{sec:watchedge_sim} introduces the WatchEDGE simulator. Sec.~\ref{sec:results} discusses the numerical analysis. Sec.~\ref{sec:conclusion} concludes the paper.

\section{Related work}\label{sec:related_work}
Federated learning, introduced by McMahan et al. \cite{mcmahan2017communication}, emerged as a privacy-preserving distributed ML method. By transmitting only model weights rather than raw data, it not only preserves user privacy but also significantly reduces communication costs. In this section, we delve into the related work concerning the integration of FL within MEC networks. 
FL operates as an encrypted distributed ML technology, allowing clients to build their own models without disclosing their data, thus safeguarding the privacy of each client's proprietary data. Subsequently, the local model parameters of each client are communicated to a central entity, where a global common model is derived \cite{zhang2021survey}. 
Numerous studies have explored FL, particularly focusing on Hierarchical Federated Learning (HFL). It consists of an extension of the FL paradigm that introduces a hierarchical intermediate structure among clients. In HFL, clients are organized into multiple levels, with each level representing a different layer of aggregation. The following studies encompass HFL strategies in terms of network and computing resource placement strategies. 

Qi et al. \cite{qi2023model} presented a systematic literature review on model aggregation in FL. The focus is on summarizing the proposed techniques and the ones currently applied for model fusion. 

Wu et al. \cite{wu2024topology} discussed the challenges and future works for applying FL to topology-specific edge networks.

Brecko et al. \cite{brecko2022federated} presented FL frameworks that are currently popular and that provide communication between clients and servers including basic models and designs of system architecture, possibilities of application in practice, privacy and security, and resource management.

Liu et al. \cite{Liu:2020} incorporated the utilization of computing resources at the network edge by exploiting HFL. As such, edge computing facilities play a crucial role in conducting intermediate aggregation of models. By strategically adjusting the frequency of partial aggregation at the edge servers and global aggregation in the cloud, they demonstrated that HFL effectively minimizes energy consumption at end devices while simultaneously reducing model training time. This hierarchical approach leverages the proximity of edge resources to devices, optimizing the efficiency of FL in distributed environments. 

Luo et al. \cite{luo2020hfel} introduced a novel Hierarchical Federated Edge Learning (HFEL) framework, where model aggregation is partially migrated to edge servers from the cloud. HFEL formulates a joint computation and communication resource allocation problem for device users to achieve global cost minimization.

Xu et al. \cite{9935309} considered cost minimization associated with joint worker aggregator placement and client assignment problems in a MEC network, optimizing client clustering, aggregator placement, and server location simultaneously.

Chen et al. \cite{9763436} described an FL approach over multi-hop wireless networks, and optimized FL over wireless networks by taking into account the heterogeneity in communication and computing resources at mesh routers and clients.

Various studies have proposed cluster-based FL mechanisms, focusing on hierarchical aggregation and convergence-bound analysis while considering factors such as energy cost, accuracy, latency, and time limitations. For instance, Wang et al. \cite{wang2021resource} show a cluster-based FL mechanism that emphasizes hierarchical aggregation; their analysis highlights that the convergence bound depends on the number of clusters and on the training epochs, with a primary focus on latency and compute time optimization.

Feng et al. \cite{9629331} focused on minimizing the energy cost while addressing constraints such as delay, local CPU-cycle frequency, power allocation, local accuracy, and subcarrier assignment. 

Meanwhile, Dinh et al. \cite{9261995} provided a convergence rate analysis, illustrating the trade-off between local computation rounds and global communication rounds in FL. They employed FEDL in wireless networks as a resource allocation optimization problem that captures the trade-off between FEDL convergence time and energy consumption of clients with heterogeneous computing and power resources.

Wu et al. \cite{10021868} introduced HiFlash, a framework that integrates deep reinforcement learning-based adaptive staleness control and heterogeneity-aware client-edge association strategy to boost the system efficiency and mitigate the staleness effect without compromising model accuracy.

Du et al. \cite{du2022bandwidth} proposed the FedMT algorithm to minimize training time through optimized client selection and routing. Their work focuses on reducing upload time per iteration within an SD-WAN topology. However, they did not explore scenarios involving intermediate computing capabilities, distinguishing our research focus.

Wang et al. \cite{10527195} proposed a minimum spanning tree (MST)-based scheduling approach that considers both resource and latency costs in Federated Learning (FL) training. Their model utilizes the shortest path to determine the optimal aggregation routing operations. Additionally, they introduce scheduling strategies \cite{wang2024poster} aimed at enhancing communication efficiency for distributed AI tasks. Their MST-based strategy dynamically determines routing paths and aggregation operations to improve the overall efficiency of the FL process.


Previous research has introduced innovative FL aggregation strategies that leverage edge computing, but a common limitation is the oversight of the edge-cloud overlay network. Specifically, the aforementioned works focus more on optimizing only up to the client-edge connection, neglecting the network scenario up to the cloud. In contrast, our work stands out by enabling model aggregation directly on selected edge nodes, achieved through the optimization of resource allocation within the edge-to-cloud overlay network.

\section{Federate Learning Resource Allocation at the Network Edge}
\label{sec:FLRS}

\subsection{Reference Network Architecture}
\label{sec:idea}

\begin{figure}[!htpb]
	\begin{subfigure}[b]{.24\textwidth}	
        \includegraphics[width=\textwidth]{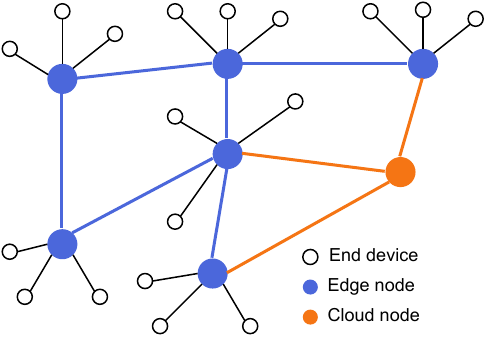}
        \caption{Example of a WatchEDGE functional topology with multiple clients end devices, edge nodes and a public cloud node.}
        \label{fig:watchedgetopo}
	\end{subfigure}
	\hfill
	\begin{subfigure}[b]{.24\textwidth}
        \includegraphics[width=\textwidth]{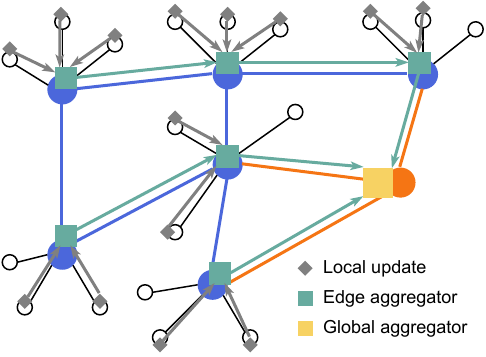}
        \caption{Aggregator overlay topology scenario. An arrow stands for an update to be routed through the link.}
        \label{fig:ov_topo}
	\end{subfigure}
\caption{Exemplary WatchEDGE topology and our proposed scenario.}
\label{fig:topo}
\vspace{-5mm}
\end{figure}

The network architecture considered in this paper is derived from the WatchEDGE project \cite{MAIER2024110248}, depicted in Fig. \ref{fig:watchedgetopo}. Within this architecture, each site comprises edge nodes. These edge sites are equipped with their own computing and networking resources, to which client end devices connect wirelessly or via physical connections. In contrast, cloud sites symbolize public cloud infrastructure and provide extensive computational capabilities and storage capacity. Inter-site connectivity is facilitated through overlay VPN tunnels, layered atop the physical access network provided by Internet Service Providers (ISPs). These VPN tunnels are centrally managed by an SD-WAN controller. End links establish connections between client devices and edge nodes, while edge links facilitate communication between edge nodes. Cloud links, on the other hand, denote connections from edge nodes to the cloud, enabling data exchange and computational offloading. This network architecture embodies edge computing, leveraging both local edge resources and remote cloud infrastructure to support diverse applications and services.

In FL model aggregation, the hierarchical two-level aggregation strategy is widely studied in the literature \cite{luo2020hfel, 9935309, wang2021resource}. Our work extends this approach \textbf{by introducing the possibility for an aggregator to report its aggregated update to another aggregator} within the WatchEDGE network, as illustrated in Fig.~\ref{fig:ov_topo}. This introduces a network of aggregators, improving aggregation efficiency and reducing overall transmission costs. By aggregating models on edge nodes before forwarding them to the cloud, we significantly reduce component capacity costs, particularly on links related to the cloud. If all models were transmitted directly to the cloud for aggregation, it would lead to substantial link capacity usage and potential congestion. However, our approach preserves the flexibility to directly report to a global aggregator in the cloud when necessary.
As shown in Fig.~\ref{fig:topo}, our model consists of a physical topology and an aggregator overlay topology. The physical topology represents the physical edge nodes and VPN links managed by an SD-WAN controller. The aggregator overlay topology (auxiliary graph) in Fig.~\ref{fig:ov_topo}, on the other hand, is used to determine potential aggregation points and find paths to the cloud, with or without bypassing aggregation nodes. While these green nodes represent potential aggregation points, the auxiliary graph allows flexibility in deciding whether to aggregate at each node. In this auxiliary graph, if a path temporarily stops at a green node (edge aggregator) all incoming models are aggregated at that node, producing a single output model. If a path uses an auxiliary link to bypass a green node, it means aggregation does not occur at that location.
To optimize this process, we first select the appropriate aggregation nodes (green nodes), which receive model updates from grey end devices and other edge nodes before performing aggregation. Once aggregation locations are determined, the next step is to establish optimal routing paths for model updates, leveraging auxiliary links in the graph (as shown in Fig. 2b) and mapping them to physical routes in the network.
This procedure introduces an additional optimization challenge, which we term the \textit{aggregator routing problem}. Our proposed aggregator overlay topology enhances flexibility beyond the traditional hierarchical two-level approach, dynamically adapting to the network's structure. To the best of our knowledge, this specific scenario has not been explored within the context of FL and MEC-based SD-WAN network. This study evaluates the cost of our aggregation method compared to conventional approaches and examines the trade-offs in FL model aggregation for both ILP and our proposed heuristic algorithm.
In this study, we employ the Training Round Failure Rate (TRFR) \cite{abdulrahman2020fedmccs} to quantify the probability that not all requests are successfully processed during an FL training phase when a set of requests is submitted to the simulator. In the FL simulator, a request refers to a training round request initiated by far-edge nodes that need to update their models to the cloud due to environmental changes, such as new data (e.g., new images) or a decline in model accuracy. The WatchEDGE infrastructure manages the initiation of these training rounds. In our simulation, requests are generated using the Application Block (AB), following a predefined inter-arrival time distribution modeled as exponential (exp). The AB is responsible for generating requests, while task management is handled by the Event Dispatcher and the Resource Allocation (RA) manager (see Section \ref{sec:watchedge_sim}). The system processes all active requests and optimizes TRFR globally by strategically selecting aggregator nodes and routing paths to enhance resource efficiency.
As a key performance metric, TRFR provides insights into system resource allocation and task execution effectiveness, allowing us to assess the impact of our proposed resource management strategies.
Additionally, we use cumulative weighted capacity as a cost metric (Section \ref{sec:ILP}), which accounts for total resource consumption. Our objective is to minimize this metric, as it directly impacts efficiency. Together, TRFR and cumulative weighted capacity play an important role in optimizing resource utilization during model training.

\subsection{Problem Statement}

We model the network as a bidirectional graph with client nodes, denoted as physical graph $G_p = (N, E)$,  where $N$ represents the set of nodes and $E$ represents the VPN tunnels (in order to be short, we call it links in the rest paragraph) in the MEC-based SD-WAN topology.  SD-WAN can dynamically create network overlays using VPN channels, enabling organizations to adapt quickly to changing network demands and configuration. In our model, the SD-WAN overlay network is considered as physical topology, and E is considered as physical links (VPN tunnels).
The nodes in N are categorized into three types: $N$ = $R \cup L \cup U$, $R$ (cloud servers), $U$ (client end devices), and $L$ (edge node aggregators). Client end devices mean the far edge in the topology, and the Edge node aggregators mean the edge node which can be considered as the aggregator nodes.  In Fig. 2, we define two types of links: Cloud Links (orange): Connecting the last edge nodes to the cloud nodes. Edge Links (blue): Connecting edge nodes to each other.
To optimize aggregation and routing, we construct an auxiliary graph where selected nodes serve as edge aggregators with computational resources, while links represent potential connections between both adjacent and non-adjacent nodes. This auxiliary graph, denoted as Ga =(Na,Ea), is fully connected, where Na represents the set of nodes that can be selected as aggregators, and Ea includes all possible links, indicating the potential to bypass certain edge nodes.
In the auxiliary graph, we first determine which nodes should act as aggregators and identify the corresponding auxiliary links to establish efficient paths to the cloud. Once the aggregator nodes and auxiliary links are properly selected in the auxiliary graph, the routing is mapped onto the physical topology to ensure efficient data transmission. The procedure for utilizing the auxiliary graph is detailed in Section III.A.
The following formula formally describes how to determine routing in both the physical and auxiliary graphs. Our goal is to aggregate the models properly using these two graphs, ensure the low TRFR and low capacity cost.

Our proposed edge-to-cloud FL model aggregation problem can be stated as follows:
\textbf{Given}: A MEC-based SD-WAN topology consisting of nodes representing clients, edge devices, and cloud servers, as well as a set of VPN channels (in order to be short, we call it links in the rest paragraph) and input requests with an inter-arrival time distribution.
\textbf{decide}: The optimal aggregation locations for nodes and establish routing and capacity assignment for all the requests in the system at that time. 
\textbf{constrained}: The resource capacity assigned to each node and link.
\textbf{objective}: Minimizing the cumulative weighted capacity cost in the ILP model (and TRFR in the heuristic algorithm, as shown in Eqn. \ref{tr} later), which together reflect the efficiency of resource utilization and the associated costs within the whole network.
We address this problem by developing an ILP formulation and a heuristic algorithm called HFEL-MESH, both described in the following sections. While many FL models use TRFR as a primary metric to evaluate system performance, it is important to note that the characteristics of the ILP approach make it unsuitable for directly assessing TRFR. TRFR is tested through simulation, relying on multiple iterations of FL behavior, whereas ILP lacks this capability. Instead, we evaluate ILP performance using cumulative weighted capacity cost, which provides a measurable method within our model’s constraints. Training rounds may fail due to (1) the inability to place an aggregator on an edge node due to insufficient resources or (2) the inability to transmit a model update through an edge or cloud link due to link capacity constraints. Since model updates are transmitted based on request demand, link capacity limitations can prevent updates, leading to an increase in TRFR. These training rounds fail are closely tied to overall resource consumption. Given that cloud links often act as bottlenecks, we assign them higher weights in the ILP to minimize their usage and prevent congestion. Meanwhile, the HFEL-MESH algorithm evaluates both TRFR and cumulative weighted capacity cost, ensuring an optimal balance between resource consumption and network performance. By efficiently distributing workloads across edge nodes, HFEL-MESH minimizes TRFR and optimizes overall resource utilization.

\begin{figure}[!htpb]
	\begin{subfigure}[b]{.24\textwidth}	
        \includegraphics[width=\textwidth]{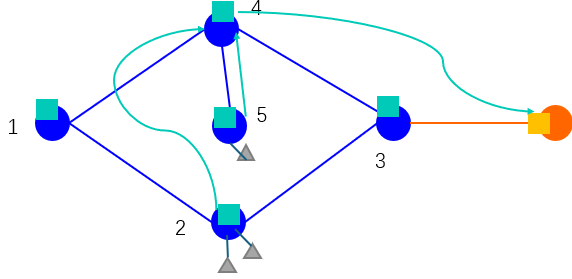}
        \caption{Route in auxiliary graph }
        \label{fig:route1}
	\end{subfigure}
	\hfill
	\begin{subfigure}[b]{.24\textwidth}
        \includegraphics[width=\textwidth]{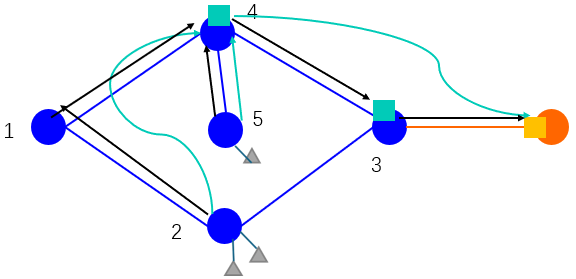}
        \caption{Route in physical graph}
        \label{fig:route2}
	\end{subfigure}
\caption{How to route the request.}
\end{figure}

Let's consider an example to illustrate the routing process. Fig. \ref{fig:route1} depicts a physical topology consisting of five nodes and five links, following the same legend as Fig. \ref{fig:topo}. In this scenario, a request requires aggregating the model from far-edge nodes. Specifically, the far-edge nodes are associated with edge node 2 and edge node 5. The aggregated model must then be transmitted to the cloud.
To determine the routing, we first analyze the auxiliary graph, which contains the same five nodes but features a fully connected topology. The chosen routing path is highlighted in green in Fig. \ref{fig:route1}. One auxiliary link starts from node 2, bypasses node 1, and reaches node 4. This implies that the model from node 2 will be sent via node 1 to node 4, where it will be aggregated. Similarly, the model from node 5 is directly transmitted to node 4 for aggregation, as indicated by the green link. At node 4, both models are aggregated before being sent to the cloud. Additionally, another green link in the auxiliary graph connects node 4 to the cloud while bypassing node 3, meaning no aggregation occurs at node 3.
Next, we determine the physical routing in the physical graph, as shown in Figure \ref{fig:route2}. Black arrows indicate the paths taken by the physical links. First, we map the auxiliary link (2,4) onto the physical graph using the path (2,1) → (1,4). The auxiliary link (5,4) directly corresponds to the physical link (5,4). Finally, we define the physical route for the auxiliary link (4, cloud) using the path (4,3) → (3, cloud).

\subsection{ILP Model Formulation}
\label{sec:ILP}
In this section, we delve into the ILP model formulation. The sets, parameters and variables are shown in Table \ref{tab:1}, Table \ref{tab:2} and Table \ref{tab:3} respectively. In the following we introduce the optimization function and constraints.

The objective function minimizes the cumulative weighted capacity utilized by nodes and links in the physical topology. This is achieved by calculating the total capacity cost of each component, weighted according to its importance, for every served request. We assign specific weights to all components, including both links and nodes, with the highest weight assigned to the cloud link to prevent congestion. Aggregation procedure cost the node capacity, and the model update cost the link capacity. Our goal is to decrease the cumulative weighted capacity cost to prevent the congestion so that to have more capacity to serve the requests and lower the TRFR as much as possible. The formal definition of the objective function is presented in Eq. \ref{ob-function}.

\begin{equation}
\min \ \ \alpha_n \cdot \eta_n +  \beta_e \cdot \eta_e \quad\forall n \in N
\label{ob-function}
\end{equation}
where $\alpha_n$ and $\beta_e$ denote the weights of node $n$ and link $e$; $\eta_n$ and $\eta_e$ measure the capacity occupied for node $n$ and link $e$. $\eta_n$ and $\eta_e$ are restricted by Eq. \ref{nn-c} and Eq. \ref{nn-e}. These two equations calculate how much capacity is occupied for each node and each link in the whole system.
In order to decrease the congestion on cloud links, their weights are much higher than that of edge links.

To optimize this process, we first use the flow constraint Eqn. \ref{flow-constraints1} to select the appropriate aggregation nodes and the proper routing in the auxiliary graph. Once aggregation locations are determined, the next step is to map the auxiliary route to physical routes in the network.
We use Eqn. \ref{f-m} to map the auxiliary link back to the physical route in the physical topology. An auxiliary link in the auxiliary graph could be consist of many physical link in the physical topology.
The constraints in our model define the routing and resource allocation in both the auxiliary and physical graphs. We denote $S^+(i)$ and $S^-(i)$ as the sets of outgoing and incoming links from node i, respectively. Eqn. \ref{flow-constraints1} enforces the flow constraint for routing in the auxiliary graph between each node pair $p \in P$, where $p$ represents the request’s source and destination nodes, denoted as $a(p)$ and $b(p)$. The source node typically represents a far-edge node, which needs to update and aggregate its model to the cloud. The destination node refers to the cloud node. The path should start at $a(p)$ and end at $b(p)$. For each auxiliary link $e \in Ea$, $i(e)$ and $j(e)$ refer to the source and destination nodes, respectively, with $j(e)$ being an aggregator. Each auxiliary link in the auxiliary graph may map to one or more physical links in the physical topology. When an auxiliary link is selected for use, its endpoint $j(e)$ is designated as an aggregator.
Eqn. \ref{f-m} enforces a constraint that if an auxiliary link is used, as indicated in Eqn. \ref{flow-constraints1}, then aggregation must occur at its destination node $j(e)$. And this node $j(e)$ will be assigned the role of an aggregator node, and it will consume computational resources accordingly.
Eqn. \ref{flow-constraints2} determines the physical route for each auxiliary link, mapping it onto the physical topology. Eqn. \ref{x-m} enforces link capacity constraints to prevent congestion. Eqn. \ref{ex} checks whether additional links should aggregate on node j in the physical topology. Eqn. \ref{m-n} and \ref{nn-c} compute node capacity costs and ensure compliance with node capacity limitations. Similarly, Eqn. \ref{m-e} and \ref{nn-e} calculate link capacity costs and enforce link capacity constraints, ensuring efficient routing and aggregation within the network.

\begin{table}[]                                     
\caption{Sets description for ILP model.}
\label{tab:1}
\begin{tabularx}{\linewidth}{|p{1cm}|X|}

\hline
\textbf{Sets}& \textbf{Description} \\
\hline

$E$ & Set of links in the physical network\\
$E_a$ & Set of links in the auxiliary graph\\
$K$ & Set of requests\\
$L$ & Set of edge nodes in the network\\
$N$ & Set of nodes in the network\\
$N_a$ & Set of nodes in the auxiliary graph\\
$P$ & Set of node pairs (client - cloud couples) for paths in topology\\
$R$ & Set of cloud servers in the network\\
$U$ & Set of user clients in topology\\

\hline
\end{tabularx}
 \\[3pt]
 \centering
\vspace{-4mm}
\end{table}

\begin{table}[]
\caption{Parameters description for ILP model.}
\label{tab:2}
\begin{tabularx}{\linewidth}{|p{0.7cm}|X|}

\hline
\textbf{Para}& \textbf{Description} \\
\hline

$C_n$ & Integer, capacity available on node $n$\\

$C_e$ & Integer, capacity available on link $e$\\

$\alpha_n$ & Float, the weight of the nodes capacity \\

$\beta_e$ & Float, the weight of the links capacity\\

$\lambda$ & Float, input request arrival rate \\

$\xi$ & Integer, Tunable hyper-parameter\\

$\sigma_n^k$ & Integer, the required node capacity of request $k \in K$\\

$\sigma_e^k$ & Integer, the required link capacity of request $k \in K$\\

\hline
\end{tabularx}
\\[3pt]
 \centering
\end{table}

\vspace{-4mm}
\begin{table}[]
\caption{Variables description for ILP model.}
\label{tab:3}
\begin{tabularx}{\linewidth}{|p{0.75cm}|X|}

\hline
\textbf{Var}& \textbf{Description} \\
\hline
$q_{e,p}^{k}$ & Binary, equals to 1 if link $e\in E_a$ is allocated for path between node pair $p \in P$ for request $k$, in the auxiliary graph\\

$x_{e',e}^{k}$ & Binary, equals to 1 if auxiliary link $e \in E_a$ selected link $e' \in E$ for connection in physical graph for request $k$\\

$t_{failed}$ & Integer, number of failed requests\\

$t_{total}$ & Integer, number of total requests\\

$z_n^k$ & Binary, equals to 1 if request $k$ used client $n \in U$ can be accepted by the model\\

$\eta_n$ & Integer, the capacity occupied on node $n \in N$\\

$\eta_e$ & Integer, the capacity occupied on link $e \in E$\\

$\mu_n^k$ & Binary, if request $k \in K$ uses node $n \in N$\\

$\mu_e^k$ & Binary, if request $k \in K$ uses link $e \in E$\\

$\Psi^{cloud}$ & Integer, cloud cost\\

$\zeta_n^k$ & Binary, if there is node $n$ used as aggregator in the physical graph\\

\hline

\end{tabularx}
\vspace{-3mm}
\end{table}

\vspace{-4mm}
\begin{footnotesize}
\begin{equation}
\setlength{\abovecaptionskip}{-7mm}
    \begin{split}
    \sum_{e \in S^+(i)} q_{e,p}^k -\sum_{e \in S^-(i)}q_{e,p}^k
    = 
    \begin{cases}
    z_{a(p)}^k \  if\ i(e)=a(p) \land j(e) \notin R\\  
    -z_{a(p)}^k \ if\ i(e)=b(p)\\
    0 \ \ \ \ others\\ 
    \end{cases}
    \\
    \quad\forall p \in P,i \in N, j \in N-R, k \in K
\label{flow-constraints1}
\end{split}
\end{equation}
\end{footnotesize}

\vspace{-4mm}
\begin{footnotesize}
\begin{equation}
q_{e,p}^{k} \leq \mu_{j(e)}^k \leq \sum_{k \in K} q_{e,p}^{k} \ \ \ \ \forall e \in E_a, k \in K, j(e) \in L \cup R, p \in P
\label{f-m}
\end{equation}
\end{footnotesize}

\vspace{-4mm}
\begin{footnotesize}
\begin{equation}
\setlength{\abovecaptionskip}{-7mm}
    \begin{split}
    \sum_{e' \in S^+(i)} x_{e',e}^{k} -\sum_{e' \in S^-(i)}x_{e',e}^{k}
    = 
    \begin{cases}
    \mu_{i(e')}^k \  if\ i(e')=a(e)\\
    -\mu_{i(e')}^k \ if\ i(e')=b(e)\\
    0 \ \ \ \ \ others\\
    \end{cases}
    \\
    \quad\forall e \in E_a,i \in N, k \in K
\label{flow-constraints2}
\end{split}
\end{equation}
\end{footnotesize}

\vspace{-4mm}
\begin{footnotesize}
\begin{equation}
x_{e',e}^{k} \leq \mu_{e'}^k \leq \sum_{k \in K} x_{e',e}^{k} \ \ \ \ \forall e' \in E, e \in E_a, k \in K
\label{x-m}
\end{equation}
\end{footnotesize}

\vspace{-4mm}
\begin{footnotesize}
\begin{equation}
\begin{split}
\mu^k_{e'(i,j)} \cup \mu^k_{e''(m,j)}  \leq \zeta_j^k  \leq \sum_{j \in L \cup R} \mu^k_{e'(i,j)} \cup \mu^k_{e''(m,j))} \ \ \ \ \
\\
\quad \forall e', e'' \in E, k \in K, i \neq m
\end{split}
\label{ex}
\end{equation}
\end{footnotesize}

\vspace{-3mm}
\begin{footnotesize}
\begin{equation}
\sum_{k \in K}(\mu_n^k \cup \zeta_n^k) \cdot \sigma_n^k  = \eta_n \ \ \ \ \forall n \in N
\label{m-n}
\end{equation}
\end{footnotesize}

\vspace{-3mm}
\begin{footnotesize}
\begin{equation}
\eta_n \leq C_n\ \ \ \ \forall n \in N
\label{nn-c}
\end{equation}
\end{footnotesize}


\vspace{-3mm}
\begin{footnotesize}
\begin{equation}
\sum_{k \in K} \mu_e^k  \cdot \sigma_e^k = \eta_e \ \ \ \ \forall e \in E
\label{m-e}
\end{equation}
\end{footnotesize}

\vspace{-3mm}
\begin{footnotesize}
\begin{equation}
\eta_e \leq C_e\ \ \ \ \forall e \in E
\label{nn-e}
\end{equation}
\end{footnotesize}


\subsection{Hierarchical Federated Edge Learning Mesh (HFEL-MESH) Algorithm} \label{sec:algorithm}

After building and evaluating the ILP formulation, we developed an heuristic algorithm called HFEL-MESH to solve the edge-to-cloud FL model aggregation problem. This type of resource allocation problems in MEC-based networks have been proved to be \textit{NP-hard} \cite{9721241}. As such, due to the scalability problem of ILP, we propose a heuristic algorithm that significantly saves computing time.

Our proposed HFEL-MESH heuristic algorithm is shown in Alg. \ref{alg:mesh}. To ensure a comprehensive comparison, we adopt the HFEL algorithm proposed by Luo et al. \cite{luo2020hfel} as a baseline for the hierarchical two-level scenario. Our HFEL-MESH algorithm builds upon HFEL, using its aggregator placement and client clustering as input to enhance comparability. In the second phase, HFEL-MESH optimizes the routing of edge aggregators by jointly selecting nodes and determining link paths through the SD-WAN.

Alg. 1 consists of two main components:
1) Optimization of resource allocation for client-to-edge association (Lines 1–12). 2) Optimization of edge aggregators overlay routing for FL model aggregation (Lines 13–31). For the first component in the algorithm (for both HFEL and HFEL-MESH), we try to associate the clients and edges. For all the possible client-edge pairs, it computes the cost for the proposed route and all involved components. The edge association procedure assigns clients to an edge node using two functions: device transfer and device exchange operations. Assuming an edge node pair (v1, v2) where a client c1 is initially associated to v1, device transfer refers to c1 being associated with v2: c1 → v2. For a pair (v1, v2) where c1 is associated to v1 and c2 to v2, device exchange operation denotes c2 → v1 and c1 → v2. To find an optimum, the algorithm iterates over all edge node pairs (v1, v2) and their respective clients until no further operation yields a lower cost per edge node pair. 

Next, we address the second component, which solves the aggregator overlay routing problem. This part is our new proposed method for HFEL-MESH. As outlined in Algorithm 1, this approach iteratively connects aggregators with the lowest routing cost. When a new request arrives, HFEL-MESH generates an auxiliary graph where each aggregator initially appears as an isolated node, corresponding to an edge node in the physical graph. After computing routing costs between aggregators and to the cloud for each pair $(a_source, a_target)$, the algorithm links the source aggregator to its potential target through a directed edge, assigning the routing cost as the edge weight.

This iterative procedure tries to locate the edge with the minimum cost, determining the first pair of aggregators to be connected. Upon connection, the overlay topology graph is checked for cycles. Then, the auxiliary graph is updated by removing the node corresponding to the source aggregator. This process is repeated for the total number of aggregators, optimizing routing for each aggregator based on cost. The computation of routing costs between aggregators takes into account the current utilization and load of both nodes and links. Eq. \ref{tr} computes the TRFR, representing the training round failure rate in a simulation run, which will be used in the HFEL-MESH algorithm. Additionally, the cost of reporting to the cloud in the algorithm is defined by Eq. \ref{eq:eta}.

\begin{algorithm}[htpb]
\algsetup{linenosize=\tiny}
\scriptsize
\caption{HFEL-MESH}\label{alg:mesh}
\KwIn{Set of clients $U$, set of aggregators $L$, and set of cloud $R$.}
\KwOut{Placed request with aggregator number and placement, client association, and aggregation topology.}
Get current network state\;
Create an auxiliary graph G'\;
$G' \gets (N_a,E_a)$ with $N_a = \emptyset, E_a = \emptyset$\;
\For{each possible node $a_\text{source} \in L$}{
    Add node $a_\text{source}$ in $G'$ \;
    \For{each possible $a_\text{target} \in L, a_\text{source} \neq a_\text{target}$}{
        $G' \gets$ add node $a_\text{target}$\;
        $G' \gets$ add edge $(a_\text{source}, a_\text{target})$ with 
        $\text{weight} = \text{cost}_{a_\text{source} \rightarrow a_\text{target}}$\;
    }
    $G' \gets$ add cloud node $c_\text{source}$\;
    $G' \gets$ add auxiliary link $(a_\text{source}, c_\text{source})$ with 
        $\text{weight} = \text{cost}_{a_\text{source} \rightarrow c}$\;
}
Create overlay graph $G''$  by connecting aggregator pairs\;
\Repeat{there has only aggregator nodes in $G'$}{
    \Repeat{overlay topology graph $G''$ has no cycle}{
        Create a set of spanning tree auxiliary edge $E_a$\;
        Try to find the smallest weight edge $e_\text{min} \gets \min_\text{weights} E_a$\;
        Extract source and target aggregator/cloud from $e_\text{min}$\;
        Connect source aggregator to its target in the overlay topology\;
        \If{cycle in overlay topology graph $G''$}
        {
            temporarily remove selected $e_\text{min}$ edge from set
            $E_a$\;
            remove connection in overlay topology graph\;
        }
    }
    remove edge $e_\text{min}$\;
    remove edge $(e_\text{min,target}, e_\text{min,source})$\;
    remove node $e_\text{min,source}$
}
Find each physical route for $E_a$ in the physical graph $G$\\
\textbf{return} overlay topology graph and physical graph\\
Calculate TRFR and cumulative weighted capacity cost\\
\end{algorithm}

\vspace{-2mm}
\begin{footnotesize}
\begin{equation}
    TRFR = \frac{t_{\text{failed}}}{t_{\text{total}}}
\label{tr}
\end{equation}
\end{footnotesize}

\begin{footnotesize}
\begin{equation}\label{eq:eta}
    \Psi^{cloud} = \frac{\mathbb{V}}{\xi  \cdot v}
\end{equation}
\end{footnotesize}Additionally, a parameter used to calculate the cost of transmitting a model through the cloud link to the cloud in the algorithm is defined by Eqn. \ref{eq:eta}. $\Psi^{cloud}$ represents the capacity cost of the cloud link and is used in Alg. 1, line 31, to compute both TRFR and cumulative weighted capacity cost. The term $\mathbb{V}$ denotes the total number of algorithm iterations, equivalent to the number of aggregators, meaning that as the number of aggregators increases, so do the algorithm's complexity and overall cost. The parameter $\xi$, a fixed integer, acts as a tunable hyper-parameter, introducing flexibility into the system. Instead of assigning a fixed cost, adjusting $\xi$ allows us to analyze the impact of varying cloud link costs \footnote{In the HFEL-MESH algorithm (Section III-D), we use a tunable hyperparameter to adjust the cloud cost. The impact of changing this parameter on cloud link utilization and system performance is demonstrated in the results.}. Increasing $\xi$ artificially lowers the cloud link reporting cost, encouraging the algorithm to prioritize cloud-based aggregation. The variable $v$ represents the current iteration number, meaning that as iterations progress, the cost associated with cloud link usage increases. To ensure consistency, we maintain the same $\Psi^{cloud}$ for both the ILP (set a proper $\sigma_e^k$ for the link cost for only one iteration test in ILP) and heuristic approaches. This process results in a graph encapsulating all feasible routing paths between aggregators and to the cloud, allowing the algorithm to facilitate local updates while selecting low-cost, low-TRFR routes.

\section{WatchEDGE simulator}
\label{sec:watchedge_sim}

\begin{figure}[h!]
    \begin{center}
    \includegraphics[width=.48\textwidth]{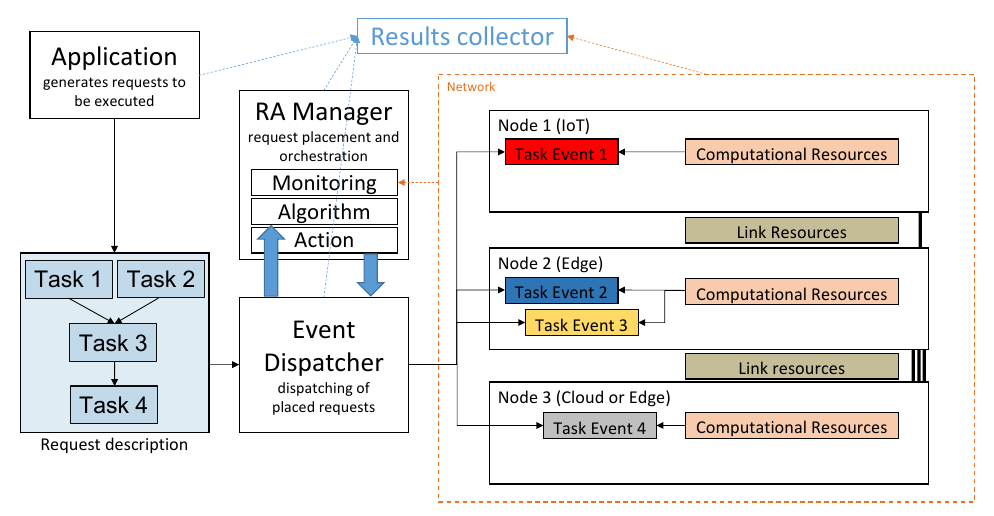}
    \end{center}
    \caption{WatchEDGE simulator architecture.}
    \label{fig:sim}
    \vspace{-8mm}
\end{figure}

To comprehend the dynamic behavior of computation and communication resources of a MEC-based SD-WAN in which FL applications are deployed, we implemented a Python-based DES. This section introduces the WatchEDGE simulator, which aims at modeling the dynamic usage of network resources where applications comprising multiple inter-dependent sub-tasks are deployed across the network. The simulator comprises five main components, as shown in Fig. \ref{fig:sim}: AB, event dispatcher, RA manager, results collector, and network. 

The AB generates requests based on the type of application being modeled. Specifically, our implementation simulates FL applications that aim at training a global model from the local updates of clients. 
These requests are generated following a predefined inter-arrival time distribution; they include information about the tasks to be deployed, their interdependence represented as an event graph, required resources, constraints, and other properties. These descriptions are then passed to the event dispatcher which collects and then forwards them to the RA manager. Upon the request arrival, the RA manager examines the request description and utilizes both the network monitoring module and the implemented algorithm, in our case this refers to the HFEL-MESH heuristic algorithm detailed in Sec.\ref{sec:algorithm}, in charge of making placement decisions.
After completing the placement task, the RA manager communicates the details of the placement decision back to the event dispatcher. The dispatcher then proceeds to deploy the requests throughout the simulated network. This deployment involves constructing an event graph, which is based on the descriptions provided in the requests.
By processing the simulation events outlined in the event graph, the request is executed, and the designated computational task interacts with the allocated resources.
The network component abstracts a real network to enable the simulation of network resource usage. For this purpose network edges and nodes host resource components that can be consumed by simulation events following an underlying resource consumption model.

In this work, nodes contain computational resources while links (assumed to be VPN tunnels of an SD-WAN) host bandwidth resources. The discussed simulator architecture shows several advantages. 

To begin with, its modular structure allows quick adaptation of the simulated resource allocation problem, e.g. adding complexity through more resource types per node or investigating a different allocation problem. For instance, the application component can be repurposed to generate other types of resource-consuming requests and the network’s behaviour solely depends on the resources and resource consumption models attached to nodes and links. Furthermore, the detached RA manager ensures an isolated implementation of each allocation method, thus guaranteeing equal conditions for different strategies.
The simulator\footnote{Apart from the descriptive information provided in this paper, we will release our source code upon acceptance of the paper in a GitLab repository, and will provide instructions on how to reproduce our numerical results.} is written in Python language (v3.9.16) and programmed to run in a Docker environment. All presented components of the simulator are implemented as Python classes following best practices of object-oriented programming whenever possible.

\section{Experiments and Results}\label{sec:results}
In this section, we outline the experiments conducted to evaluate our proposed algorithms. Initially, we introduce the simulation settings encompassing the used network topologies, computing and network resources, and the specific use case under consideration for this study. Subsequently, we present a comparative analysis between our proposed ILP and heuristic algorithm implementations, comparing them with a state-of-the-art algorithm proposed by Luo et al. \cite{luo2020hfel}.

\subsection{Simulation settings}

To evaluate the performance of the proposed HFEL-MESH algorithms, we integrated them into the WatchEDGE simulator as discussed in Section \ref{sec:watchedge_sim}. This simulator facilitates in-depth simulations of edge networks and computing infrastructure leveraging SD-WAN and MEC technologies. Experimental assessments were carried out on a machine powered by an Intel(R) Core(TM) i7-9700 CPU.

As stated in the Introduction section, our use case revolves around environmental surveillance of wildlife. Specifically, we consider a set of end devices, acting as clients, capable of capturing images from the environment, akin to stationary smart cameras. Each client possesses the capability to gather images, constructing a local dataset, and run ML training aimed at identifying specific events, such as the presence of certain wild animals or the initiation of a wildfire. Within our context, we utilize this use case to configure the input settings of our simulator and conduct experiments. It is important to note that our focus lies on network resource allocation for FL applications rather than solely optimizing the accuracy of FL-based image recognition models. 

\begin{figure}[!htpb]
\vspace{-5mm}
	\begin{subfigure}[b]{.24\textwidth}	
		\includegraphics[width=\textwidth]{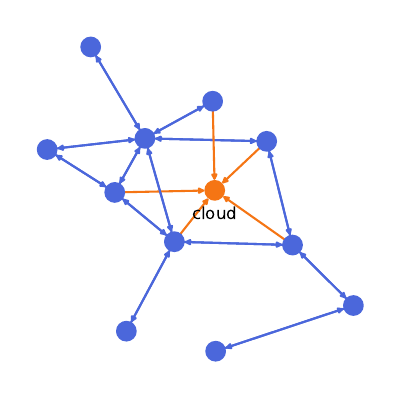}
		\caption{Medium topology with $11$ edge nodes.}
		\label{fig:medium}
	\end{subfigure}
	\hfill
	\begin{subfigure}[b]{.24\textwidth}
		\includegraphics[width=\textwidth]{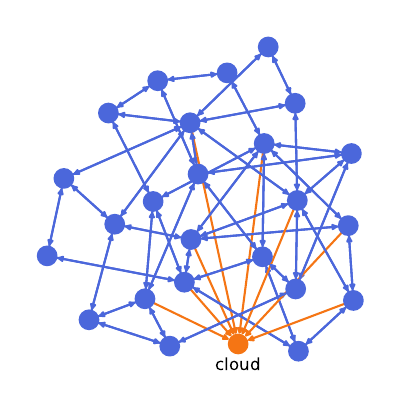}
		\caption{Large topology with $24$ edge nodes.}
		\label{fig:large}
	\end{subfigure}
\caption{Network topologies used for experiments. Blue nodes are edge nodes, and orange nodes represent a public cloud.}
\label{fig:topos}
\vspace{-3mm}
\end{figure}

We conducted evaluations on two network topologies, labeled as medium and large, as shown in Fig. \ref{fig:topos}, which were adapted from the work of Xiang et al. \cite{xiang2021dataset}. The medium topology comprises 11 edge nodes, each supporting 20 end devices per edge node, while the large topology comprises 24 edge nodes, each with 10 end devices. Additionally, each topology includes a cloud node to facilitate model aggregation.

We simulate an FL application in the simulator to generate training rounds for requests that trigger training processes on client end devices and subsequent model aggregation to both edge nodes and the cloud. The implemented algorithms are responsible for optimizing the distribution of model aggregation across the network, potentially leveraging techniques such as intermediate model aggregation through edge aggregators. Notably, the FL application generates independent training round requests, simulating multiple model owners seeking to train their respective models.

Our primary focus is not on comparing the latency of our approach with standard Federated Learning (FL) procedures. Instead, we aim to simulate the specialized HFEL-MESH aggregation, routing, and resource allocation mechanisms.  
To accurately simulate the training process, we deviate from conventional approaches of expressing training time in terms of CPU cycles. Instead, we rely on experimental results reported by Rajagopal et al. \cite{Rajagopal:2021}. For simulating model aggregation, we adopt a CPU-cycle-based computation latency model proposed by Liu et al. \cite{Liu:2020}. The communication latency to the cloud is calculated as the number of CPU cycles required to execute one request, multiplied by the dataset size and divided by the CPU cycle frequency allocated to the training task. Moreover, we specify the capacities assigned to links and nodes, as outlined in Table~\ref{tab:sim_param}.This approach allows us to assess HFEL-MESH’s performance in edge computing environments, with a particular emphasis on resource management and aggregation. By concentrating on these elements, we provide a more focused analysis, rather than emphasizing the broader FL training process. In our model, the edge aggregation process consumes resources at the edge aggregator, corresponding to specific nodes within the network topology. During this process, multiple models (sets of parameters) are received as inputs, but only a single aggregated model (set of parameters) is produced. Consequently, this reduces the amount of link resources required for transmission after aggregation.

The inter-arrival time between consecutive requests is modeled with an exponential distribution. The capacities requested at nodes and links during simulation events are generated from a uniform distribution \cite{liu2020dynamic}\cite{burd1996processor} as specified in Table~\ref{tab:sim_param}. The clients participating in each training round represent a random subset of all clients. The number of clients associated with each edge node is denoted by the symbol $\sharp$clients. To simulate the training at clients, the FL application randomly selects an image dataset size and a model architecture from a predefined set. The training times for a single image are detailed in Table~\ref{tab:sim_param}. Additionally, we consider the number of weights for each model architecture in the computation of the aggregation latency.

\begin{table}[]                                     
\caption{Overview of simulation parameters.}
\begin{tabularx}{\linewidth}{|p{3cm}|X|}

\hline
\textbf{Network resources}&  \\
\hline
client & $1$ computing unit\\ 
edge node & $200$ computing units\\ 
cloud node & $4000$ computing units\\ 
end link & \SI{200}{Mbps}\\ 
edge link & \SI{2000}{Mbps}\\ 
cloud link & \SI{4000}{Mbps}\\
\hline
Training round request & \\
\hline
requested node capacity & $[1, 8]~\text{cores}$\\
requested link capacity & $[20, 40]~\text{Mbps}$\\
client number & $[\frac{\text{\# clients}}{4}, \frac{\text{\# clients}}{2}]$\\
dataset size & $[59, 118]~\text{images}$\\ 
\hline
Model architectures & single-image training time, \#weights\\
\hline
Squeezenet & \SI{26.4}{ms}, \SI{421098}{weights} \cite{iandola2016squeezenet} \\
MobileNetV2 & \SI{38.4}{ms}, \SI{3400000}{weights} \cite{zhu2018mobilenetv2}\\ 
MNas & \SI{35.7}{ms}, \SI{3900000}{weights} \cite{tan2019mnasnet}\\
GoogleNet & \SI{35.9}{ms}, \SI{6797700}{weights} \cite{szegedy2015going}\\
Res18 & \SI{21.8}{ms}, \SI{11689512}{weights} \cite{Contributors:2023a}\\
Res50 & \SI{77.8}{ms}, \SI{25557032}{weights} \cite{Contributors:2023a}\\
\hline
\end{tabularx}
 \\[3pt]
 \centering
 \label{tab:sim_param}
\vspace{-8mm}
\end{table}


\subsection{Evaluating ILP, Heuristic and Baseline Algorithms on Medium-size Topology}

In this section, we evaluate the performance of the ILP, HFEL-MESH and the baseline HFEL algorithm, proposed by Luo et al. \cite{luo2020hfel}, in terms of Mean Utilization Ratio (MUR) and objective function. We define the MUR for links and nodes respectively as follows:

\begin{footnotesize}
\begin{equation}
MUR_{link} = \frac{\eta_e}{C_e}\ \ \ \ \forall e \in E
\label{u-1}
\end{equation}
\end{footnotesize}

\begin{footnotesize}
\begin{equation}
MUR_{node} = \frac{\eta_n}{C_n} \ \ \ \ \forall n \in N
\label{u-2}
\end{equation}
\end{footnotesize}where $\eta_e$ and $\eta_n$ are the used capacity on links and nodes, respectively, while $C_e$ and $C_n$ denote total available capacity on links and nodes.

The objective function aims at minimizing the cumulative weighted capacity, reflecting the effectiveness of capacity allocation. A lower objective function signifies a better distribution of capacity, indicating a balanced aggregation of FL data flow and a reduced cost on less critical nodes (e.g. congested nodes or links). In essence, achieving a lower objective function demonstrates optimal capacity utilization and efficient resource allocation across the network.

\begin{figure}[!htpb]
	\begin{subfigure}[b]{.32\textwidth}	
		\includegraphics[width=\textwidth]{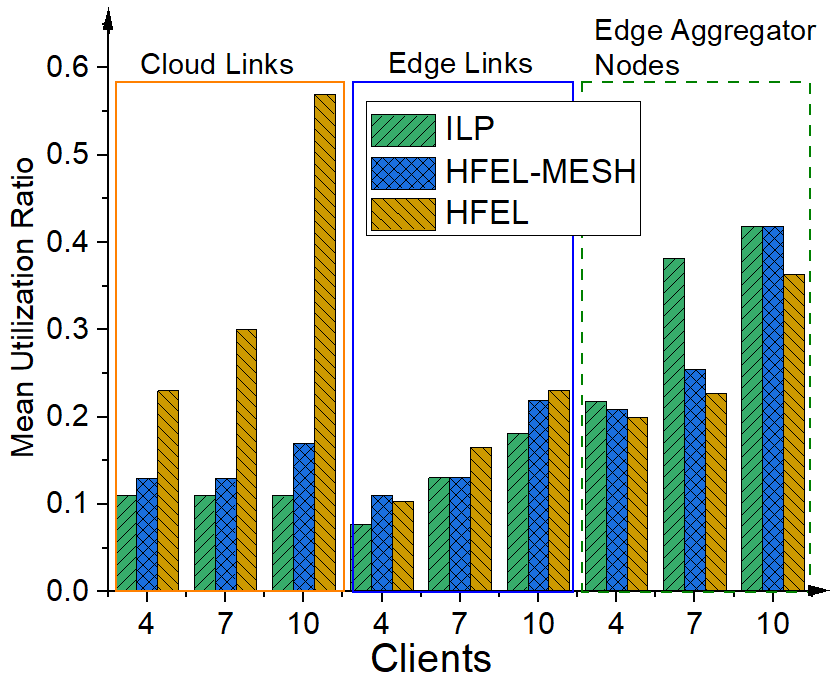}
		\caption{Utilization ratio}
		\label{fig:ilp1}
	\end{subfigure}
	\hfill
	\begin{subfigure}[b]{.16\textwidth}
		\includegraphics[width=\textwidth]{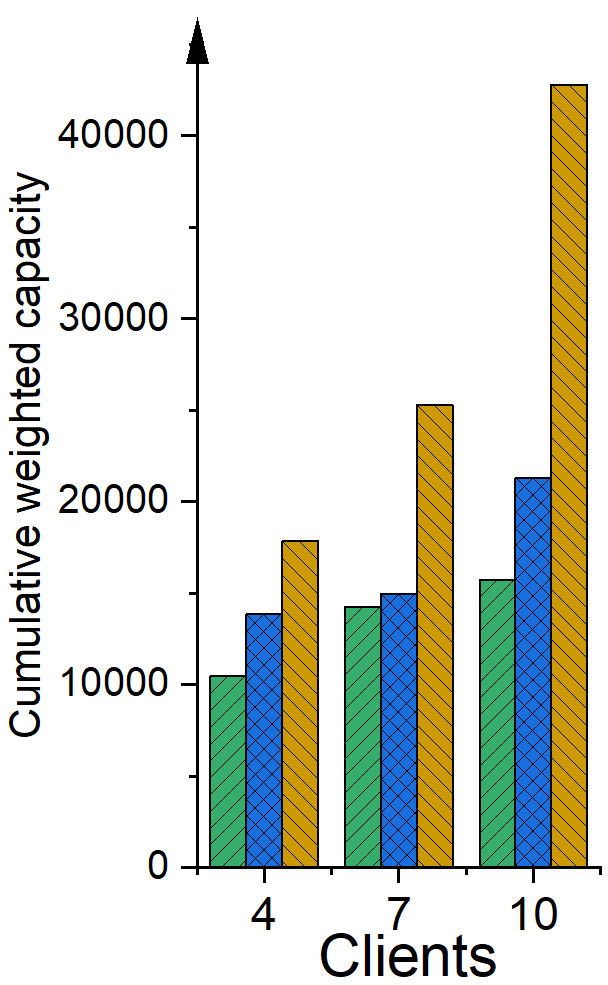}
		\caption{Objective function}
		\label{fig:ilp2}
	\end{subfigure}
\caption{Comparative analysis between ILP, HFEL, and HFEL-MESH.}
\label{fig:ILP}
\vspace{-5mm}
\end{figure}

We conducted performance evaluations using the medium-size topology illustrated in Fig. \ref{fig:medium}, accommodating a maximum of 10 requests due to scalability constraints with ILP. Each request involved allocating training rounds to different clients per edge node. In our experiments, we considered from 4 to 10 clients per request.  
Our assessment focused on the algorithms' effectiveness in minimizing the MUR of three key components: Cloud link, edge link, and edge aggregator nodes, as mentioned before.
Fig.~\ref{fig:ilp1} shows the results of the MUR by varying the number of clients. 

Our analysis reveals that the mean utilization ratio of cloud links for HFEL-MESH is between 15\% - 23.5\% higher than that of ILP, while being between 23.5\% - 43\% lower than that of the baseline HFEL algorithm. These findings underscore the effectiveness of our HFEL-MESH algorithm in substantially reducing utilization and congestion on the cloud links compared to the baseline HFEL approach. Furthermore, HFEL-MESH achieves a performance level closely aligned with ILP, as expected.


However, HFEL-MESH exhibits a maximum 30\% optimality gap in edge aggregator node utilization compared to ILP, and 5\% lower utilization compared to the HFEL baseline algorithm. This discrepancy can be attributed to the distinct objectives pursued by these algorithms. With HFEL-MESH, multiple aggregations occur on the edge nodes, resulting in higher edge node utilization compared to HFEL. In particular, for scenarios involving requests with 7 clients, HFEL-MESH achieves a 15\% higher utilization of cloud links compared to ILP, while ILP incurs a 52\% higher utilization of edge aggregator nodes compared to HFEL-MESH. This disparity arises from the higher weight assigned to cloud links in the ILP's objective function, which drives it to prioritize reducing cloud link utilization, even at the cost of increased edge aggregator node utilization. In contrast, HFEL-MESH achieves a more balanced utilization across these components. The ILP is designed to minimize cloud link usage to prevent cloud congestion, while the heuristic approach focuses more on reducing the TRFR and makes the model more balanced. HFEL aims for a more balanced outcome overall. Although ILP is expected to deliver better performance, HFEL-MESH, while not optimal, offers a practical balance between resource utilization and TRFR.

Furthermore, as depicted in Fig.~\ref{fig:ilp2}, we observe that the objective function of HFEL-MESH is at most 32\% higher than ILP, yet significantly lower than the baseline algorithm HFEL. This reaffirms the superior performance of HFEL-MESH over HFEL, with the performance gap widening as the complexity of requests increases.


\begin{figure}[h!]
    \begin{center}
    \includegraphics[width=.4\textwidth]{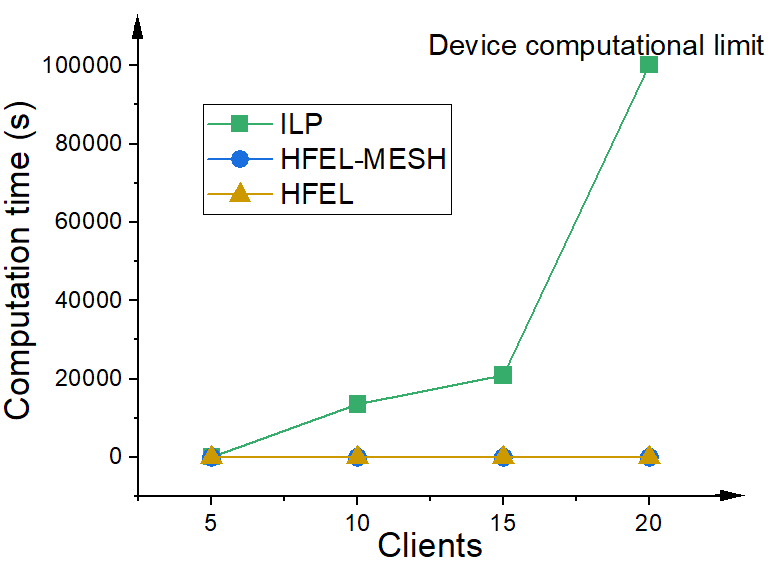}
    \end{center}
    \caption{Computational time.}
    \label{fig:time}
    \vspace{-5mm}
\end{figure}
Then, we examine the complexity of three different methods. Given the scalability challenges associated with the ILP approach, we tested the computational time of the system using a single training request with varying numbers of clients. As illustrated in Fig. \ref{fig:time}, the computational time for ILP increases nearly exponentially as the number of clients grows. When the training request involves 20 clients, the ILP method reaches the computational limit of our device. In contrast, HFEL and HFEL-MESH exhibit nearly constant computational times across all scenarios, as shown in Fig. \ref{fig:time}: even as the number of clients increases, the computational time for both HFEL and HFEL-MESH remains relatively stable.

\subsection{Evaluating Heuristic and Baseline Algorithms on Large-size Topology}

In this section, we evaluate the performance of the HFEL-MESH heuristic and the baseline HFEL algorithms in terms of MUR, as defined by Eq. \ref{u-1} and Eq. \ref{u-2}, and Training Rounds Failure Rate (TRFR) defined by Eq. \ref{tr}.

We conducted performance evaluations using the medium-size and large-size topology shown in Fig. \ref{fig:medium} and Fig. \ref{fig:large}, respectively.

\begin{figure}[h!]
        \vspace{-4mm}
	\begin{subfigure}[b]{.48\textwidth}
		\includegraphics[width=\textwidth]{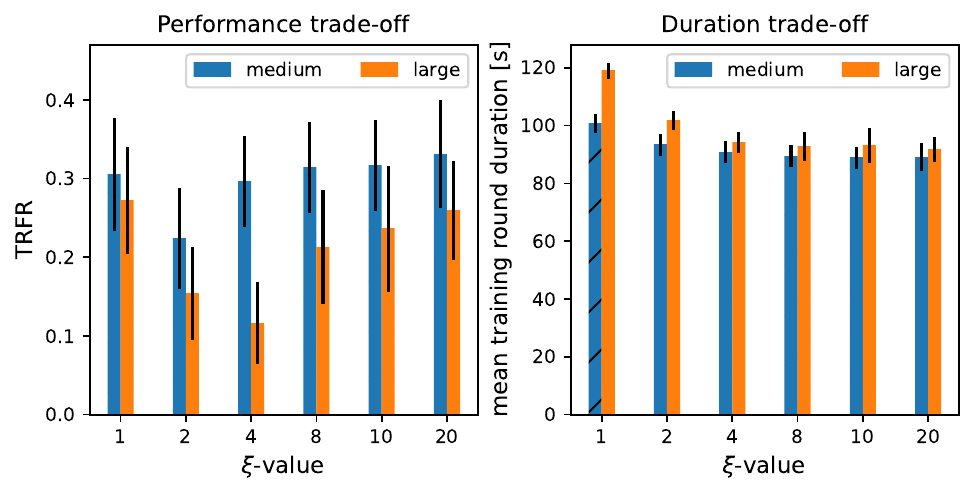}
		\caption{TRFR of HFEL-MESH (left) and training round duration (right) for different $\xi$-values \cite{burd1996processor}.}
		\label{fig:trfr_dur}
	\end{subfigure}
	\vfill
		\begin{subfigure}[b]{.48\textwidth}	
		\includegraphics[width=\textwidth]{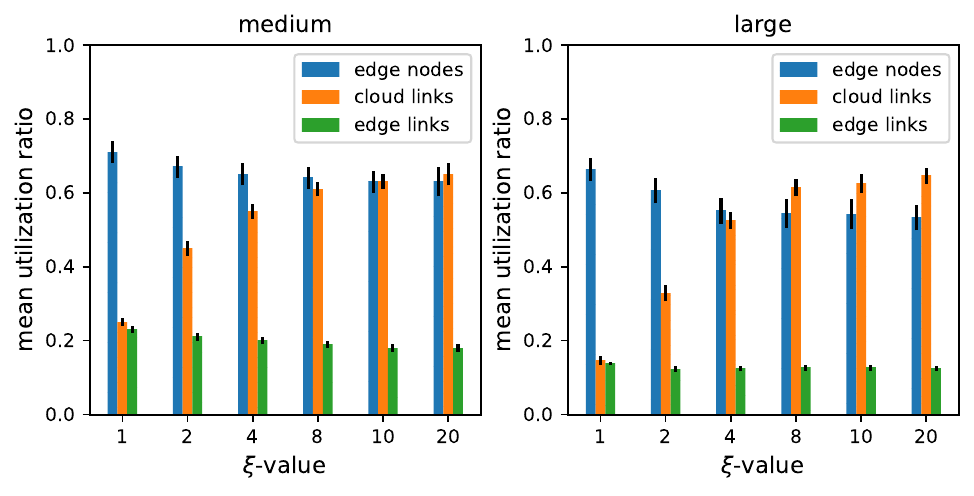}
		\caption{Mean utilization ratio of HFEL-MESH for different $\xi$-values.}
		\label{fig:mean_ut_hmesh}
	\end{subfigure}
\caption{Demonstration of edge-cloud trade-off by tuning $\xi$-parameter.}
\label{fig:reqrate_sfc}
\end{figure}

Fig. \ref{fig:trfr_dur} illustrates the outcomes of the TRFR and the mean training round duration of the HFEL-MESH algorithm in response to variations of the hyper-parameter $\xi$. This parameter, $\xi$, serves as a tunable factor enabling adjustments to the algorithm's inclination towards reducing the number of edge aggregators or vice versa. A lower value of $\xi$ favors aggregation on edge nodes (towards the edge nodes directly connected to the clients), whereas a higher value prioritizes aggregation on the cloud node (towards a total aggregation in the cloud).
As depicted in Fig. \ref{fig:trfr_dur}, we can notice that leveraging edge nodes for aggregations provides a distinct advantage, resulting in approximately an 11\% of TRFR in the large topology.

Interestingly, when $\xi$=1, there is an observed increase in the TRFR for both network topologies. This suggests that an excessive emphasis on (far) edge nodes directly connected to the clients becomes counterproductive, and leads to longer training round durations as shown in Fig.~\ref{fig:trfr_dur} (right). In other words, the algorithm prioritizes aggregating every local model update on the nearest edge node, which may result in routing detours and unnecessary capacity waste. This wastage contributes to an increase in TRFR and in a longer training round durations.


The higher TRFR observed for the medium topology can be attributed to its increased utilization ratio, as depicted in Fig.~\ref{fig:mean_ut_hmesh} (left). Conversely, the large topology, characterized by a lower TRFR, exhibits a correspondingly lower utilization ratio compared to the medium topology, as shown in Fig.~\ref{fig:mean_ut_hmesh} (right). It is interesting to note the increase in the cloud links usage with growing $\xi$ for both topologies, while edge nodes utilization gradually decreases, albeit at a slower rate. This highlights the advantage of an aggregator overlay topology: the additional occupation of edge nodes, associated with the presence of edge aggregators, results in a smaller increase in utilization compared to the reduction in cloud links utilization.


Considering the duration of a training round, it is noteworthy from Fig.~\ref{fig:trfr_dur} and Fig.~\ref{fig:mean_ut_hmesh} that increasing edge nodes utilization results in longer training round durations. This observation suggests that although higher load distribution and lower TRFR are achieved, they are counteracted by increased request durations. The rationale behind these findings lies in the prioritization of load distribution through edge aggregators during the creation of the aggregator overlay network, without considering request duration as a constraint. This leads to longer training round durations due to the additional aggregation steps introduced.


\begin{figure}[h!]
     \vspace{-5mm}
	\begin{center}
    \includegraphics[width=.48\textwidth]{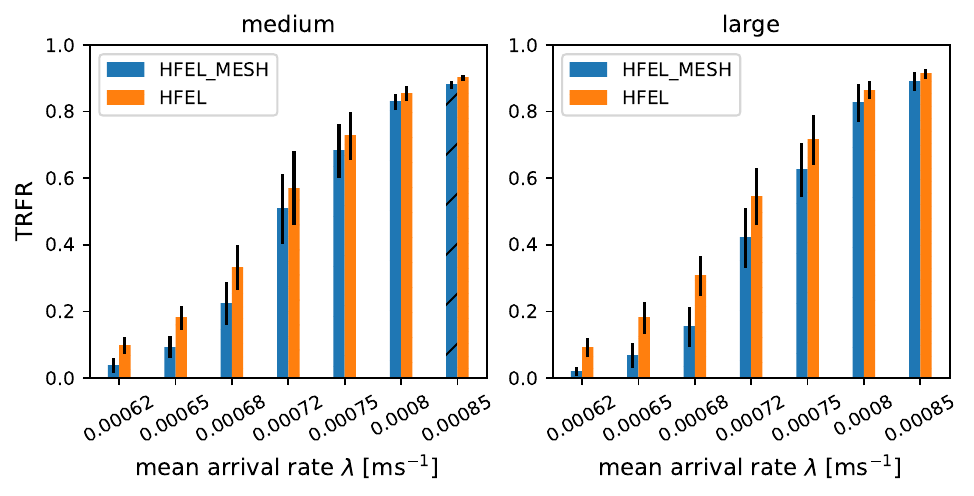}
    \end{center}
    \caption{Training round failure rate over arrival rate $\lambda$. The system reacts sensibly to a change in the arrival rate.}
    \label{fig:reqrate_sfc_trfr}
\end{figure}

Evaluating different training round arrival rates $\lambda$ provides insight into the system's capacity to handle requests. Fig. \ref{fig:reqrate_sfc_trfr} reports a TRFR below $\SI{10}{\%}$ for $\lambda = 0.00062\,\text{ms}^{-1}$ and nearly $\SI{90}{\%}$ for $\lambda = 0.00085\,\text{ms}^{-1}$ across different topologies. These $\lambda$ values result in average inter-arrival times between $\SI{1.2}{\second}$ and $\SI{1.6}{\second}$ on average, emphasizing the delicate balance that needs to be maintained by the algorithms. This fragility stems from the fact that a training round has a significantly longer lifespan compared to the arrival rate, making the system susceptible to saturation with just a few additional arriving requests.

HFEL-MESH consistently achieves a lower TRFR over topologies and arrival rates on average. For instance, the difference reaches up to $15\%$ for the large topology, underscoring the value added by routing and aggregation between aggregators. Combining this with the reported resource utilization in Fig. ~\ref{fig:mean_ut_hmesh}, we can state that HFEL-MESH optimizes computing resource utilization by increasing the usage of edge nodes while achieving a lower TRFR. Consequently, HFEL-MESH operates more reliably and efficiently. However, the difference in TRFR reported is smaller for the medium topology and diminishes further with an increased arrival rate. This is attributed to the lower availability of edge computing nodes, which restricts the aggregator routing capabilities of HFEL-MESH. Regarding the mean request placement time, we observe that even in the large topology the overhead of routing between aggregators is acceptable with $\SI{0.55}{\second}$ for HFEL-MESH compared to $\SI{0.5}{\second}$ for HFEL ($\lambda = 0.00068\,\text{ms}^{-1}$).

This research underpins the WatchEDGE project, which aims to investigate an advanced edge computing architecture distributed across multiple geographically distant sites. Our primary focus is on orchestrating limited network resources while reducing cloud link utilization. The results show that HFEL-MESH achieves balanced resource utilization and significantly reduces cloud link usage compared to the HFEL baseline. While HFEL-MESH exhibits a slight performance gap when compared to the optimal ILP solution, it still proves to be a more efficient alternative, particularly in lowering the Training Round Failure Rate (TRFR) compared to the HEFL algorithm. This suggests that HFEL-MESH enhances FL performance relative to existing methods.
The topology employed in this research is based on a real-world scenario, with plans to develop a custom topology in future work. Our approach to cloud utilization has demonstrated its effectiveness, further validating the potential of HFEL-MESH in practical applications. In conclusion, online resource-aware aggregation proves to be both advantageous and feasible for optimizing network resource utilization and reducing TRFR in networks with sub-optimal cloud connectivity and limited edge node resources. However, the presented analysis disregards other relevant aspects of edge computing, i.e. edge node reliability and trust \cite{carvalho2021edge}, which remain beyond the scope of this paper.

\section{Conclusions}\label{sec:conclusion}

In this paper, we investigate the integration of FL into the WatchEDGE network infrastructure, a specialized MEC architecture powered by SD-WAN and AI-driven end devices. Through the utilization of ILP-based and HFEL-MESH algorithms, we assess the impact of FL training on network resource usage. Our findings reveal a discernible trade-off between computation at the edge and communication to the cloud, favoring edge computing for reduced TRFR and enhanced network capacity utilization. Furthermore, we introduce a novel online aggregation technique that introduces routing capabilities between edge aggregators. Our performance analysis demonstrates the superiority of this technique over state-of-the-art aggregation methods in adapting to the network's resource characteristics. This novel approach yields a significant reduction in TRFR by up to 15\%, concurrently mitigating the cloud communication overhead induced by FL. Looking ahead, our future research endeavors will delve into the incorporation of additional FL-specific metrics, such as energy cost and node reliability, into our aggregator routing strategy. By exploring further implications of FL in MEC networks, we aim to provide comprehensive insights into the optimization of FL-enabled network infrastructures.

\section*{Acknowledgement}
\footnotesize
This work was supported by the European Union - Next Generation EU under the Italian National Recovery and Resilience Plan (NRRP), Mission 4, Component 2, Investment 1.3, CUP D43C22003080001, partnership on “Telecommunications of the Future” (PE00000001 - program “RESTART”).


\bibliographystyle{ieeetr}
{\footnotesize
\bibliography{sample}}



\end{document}